# Upgrade of the pedagogic and popular science tool for holography to a colour version

A. Escarguel
alexandre.escarguel@univ-amu.fr
*Laboratoire PIIM, Aix Marseille Univ, CNRS, UMR 7345, avenue escadrille Normandie-Niemen, Marseille, 13013, France*
.

**Abstract**

In 2012, a pedagogic tool for monochromatic holography was realized for pedagogic and popular science purposes [Th. Voslion and A. Escarguel, Eur. J. Phys **33**, 1803 (2012)]. Following its success, we decided to upgrade it to make larger colour holograms and new pedagogic experiments. The resulting kit includes all the necessary equipment to produce 4" x 5" colour holograms with a simple optical assembly and with an excellent vibration tolerance. The resulting holograms, in colour and bigger, are much more spectacular for science outreach purposes. For teaching purposes, some of the existing experiments have been upgraded, and new ones have been developed for university students and continuing education of teachers: colour reflection Denisyuk holograms, single shot transmission/reflection holograms, angular and wavelength multiplexing, holographic diffraction gratings with improved setup.

1. **Introduction**

The reproduction and the capture of image has always been a fascinating challenge for mankind. First by using painting: from the paints in the Lascaux cavern in France aged of 18 000 years, through the beautiful paintings of the classical period illustrated for example by Leonardo da Vinci, human approached the perfection in the reproduction of Nature. In the 19$^{th}$ century, photography was invented and constantly improved to reproduce the reality in black and white, and later in colour (Lippmann, autochroms and additive/subtractive synthesis techniques). The last phase of this development was the possibility to catch the relief of a real scene. For this, two shifted images must be sent to each of our eyes. Several techniques were developed with (anaglyph), or without (lenticular techniques, holography) glasses.

Holography has a particular place in this history, as it is the only way to record the "whole" relief of a scene. Indeed, other processes record the 3D effect for one or a few points of view. Holography records a huge set of points of view of the scene, and for some holograms you can have a 3D effect over 180°. This is due to the fact that interference is a non-local effect. Indeed, each point of the object illuminated by the coherent wave diffuses the information over a wide part of the photosensitive plate, giving rise to an interference pattern which contains the 3D information. A hologram has a strong redundancy of information, which is why it is possible to see the whole object with only one part of the plate. Whoever looks at a hologram thinks that holography is a "kind of magic", even though the underlying physics of interference of coherent light is well understood nowadays. Then, it is a very motivating tool to interest students and to teach them fundamental/applied optics. For popular science purposes, the realization of a live hologram has a strong impact on the audience, which allows opening discussions with people and thereby sharing the knowledge created in the research laboratories. In this way, this is a tool to fight against obscurantism and the "black box syndrome".

This paper deals with the upgrade of a pedagogic tool for monochromatic holography developed a few years ago [1, 2, 3, 4, 5, 6, 7]. The addition of a second green laser mixed with the red one allows realizing colour holograms, and new/upgraded experiments are proposed for university students and continuing education [8, 9]. The next part describes in detail the kit while the third one presents the experiments realized with it.

**2. Description of the kit for colour holography**

The realization of a hologram needs at least one coherent light source whose beam is divided in two parts, the object and the reference beams. The first beam is directed onto the object and keeps a footprint of the wavefront deformation induced by its relief. The second beam is directed on the recording medium. Its interference with the object beam allows recording the phase deformation induced by the object, and then its relief. For diffusing objects, this information spreads over the hole hologram [10].

The main element of the kit for colour holography is an optical setup fixed on a small (400 x 400 mm) breadboard with sorbothan isolating foot fixed in the bottom of the case (figure 1). It has been optimized to be simple, robust and to allow easy optical alignment when necessary. This setup is made of a 20 mW 632.8 nm HeNe laser, a 20 mW 532 nm DPSS laser (class IIIb), a dichroïc mirror to merge the collimated beams, a rotating polarizing filter in front of the green beam to control the red/green intensity ratio, and a convex lens placed in the path of the two mixed beams, just before an exit hole made in the case. A zero aperture iris is placed just before the lens so that it is possible to block the lasers beams. To decrease the hazard associated with collimated laser beams, only diverging beams are accessible outside the case. The HeNe laser has a coherence length around 150 mm, while the green laser is monomode with a much larger coherence length. Then, for Denysiuk single beams colour holograms, the maximum distance $d$ between the holographic plate and the object is limited by the red laser coherence, namely around 75 mm.

The mirror and the dichroïc mirror are fixed on kinematic mounts with manual adjusters for easy optical alignments. As in the previous pedagogic kit for holography, all the necessary equipment needed to develop the holographic plates is inside the case. This comprises the developing products (developer, bleach and wetting agent) and the corresponding tools (thermometer to control the temperature of the developer, syringe, measuring beaker, hair dryer to dry the plates after development…) and a few 4" x 5" Ultimate 08 colour plates [11].

Two plate/object supports have been developed. The first one (PO1) is an upgrade of the support proposed in the monochromatic pedagogic case. It has been redesigned and optimized for 4" x 5" plates (figure 2): the reflecting prism has been replaced by a cheaper and bigger front surface mirror to reflect the laser beams onto a large part of the holographic plate. For better stability, three 3 mm diameter metallic balls have been fixed under the support, so that it stands on three points. The same principle has been used for the photosensitive holographic plate, which is put on three columns with the same kind of balls at their ends. The two laser beams are mixed with the dichroïc mirror, expanded with the lens, and come horizontally onto the PO1 support. They are deviated vertically by the 45° mirror to cross the transparent photosensitive plate and to be back-diffused by the object which is simply put on the top of the plate. This allows taking advantage of gravity to avoid any relative movement during the realization of the hologram, so that holograms can be made even in noisy environments like during a course in a full amphitheater at the university or during a science outreach session with kids. This robustness is due to the fact that in holography, one needs to prevent any sub-micrometric perturbating vibrations only after the separation of the laser beam into the reference and the object beams [10]. In our case, as described in the previous article [1], this implies to avoid any vibration of the object with respect to the plate on which it is placed. As in the first kit, the plate/object holder is simply placed far enough from the case so that the combined red/green diverging beams deviated by the 45° mirror lighten the largest surface of the photosensitive plate. Typical exposure times are around a few seconds. This is sufficiently long to precisely manually expose the plate to the lasers by simply removing a black obturator placed on its way. And this is short enough to avoid possible perturbating vibrations.

A second plate/object support (PO2) is also included in the kit to realize single shot reflection/transmission holograms (figure 2). It allows fixing vertically the holographic plate onto the support with the aid of four screws and plastic washers. The objects can be placed behind or in front of the plate and are stabilized onto the support with some small pieces of modeling clays. A small mirror can be added on the side to improve the brightness of the transmission hologram by deviating a part of the beam on the object side in front of the plate.

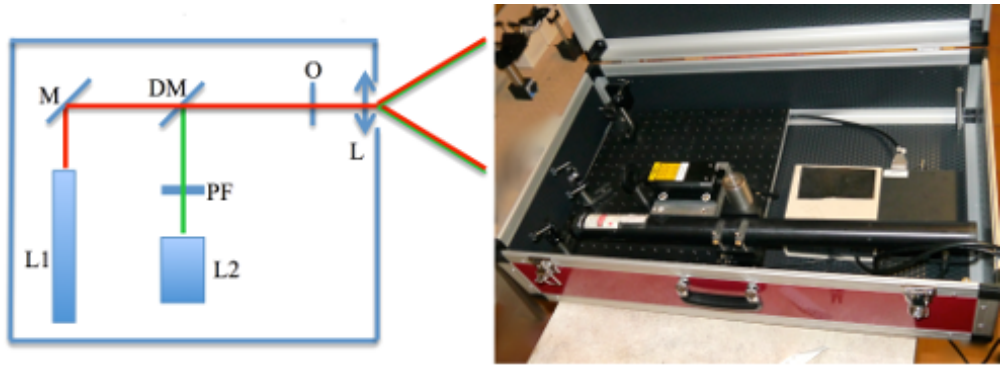

**Figure 1** Diagram of the optical setup inside the case (left) and its photography (right). L1: 20 mW 632.8 nm HeNe laser ; L2: 20 mW 532 mn DPSS laser ; PF: polarizing filter ; DM: dichroïc mirror ; M : front surface mirror ; O : obturator ; L : lens. On the photography, the exit port and the lens L are located on the upper left side of the case. A small case contains the supplementary material and is placed on the top of the lasers power supplies.

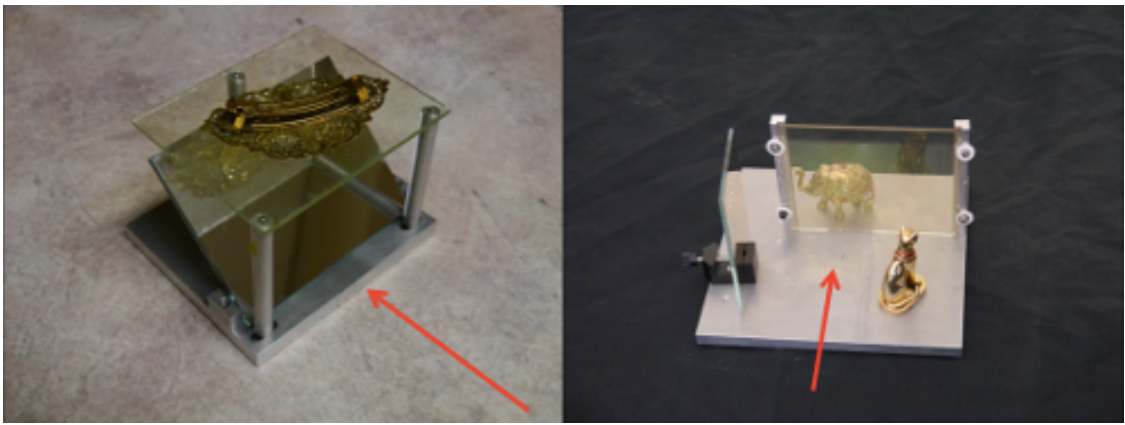

**Figure 2** Plate/object supports PO1 (left) and PO2 (right). The arrows indicate the orientation of the diverging laser beams. The plate/object supports are placed far enough from the lasers so that the maximum surface of the 4"x5" plates are exposed. PO2 is used for single shot reflection/transmission holograms: a reflection hologram of the elephant and a transmission hologram of the cat are produced simultaneously.

## 3. Experiments for popular science and pedagogy

### 3.1. Presentation

The kit for colour holography allows the realization of the following experiments:

- Monochromatic and colour Denysiuk reflection holograms,
- Single shot transmission/reflection holograms,
- Diffraction grating with groove density control (emission spectroscopy),
- Wavelength and angle multiplexing, illustrating data storage (Holographic Versatile Disc).

Moreover, as with the kit for monochromatic holography, it is possible to realize experiments on double exposure holographic interferometry (mechanical stress diagnostics) and notch filters (Raman spectroscopy). For more information on these experiments, the reader can refer to the previous publication [1]. They are enriched by the presence of two lasers in the new version of the kit. For example, it is possible to realize a double notch filter by using a front surface mirror exposed to the red and green lasers at the same time, with two reflection spectral bands centered on the lasers wavelengths.

### 3.2. Colour Denysiuk reflection holograms

For science outreach purposes, the main advantage of the kit is to allow the realization of spectacular 4"x5" Denysiuk colour holograms [12]. The redesigned plate/object support PO1 releases the experimentalist from unwanted disturbing vibration. The device was used for the successful realization of holograms in noisy environments as an amphitheater full of students. For perfect reproducibility of a wide range of colour shades, one need to use at least three lasers with wavelengths corresponding to three independent primary colours [13]. However, the realism of « colour » holograms realized with red and green lasers is already very impressive, even though it is not possible to obtain pure white and purple/blue shades (figure 3). When they are made during science outreach events, the impact on the audience of such « two colours » 4"x5" holograms is very strong, much more than small monochromatic ones. Then, the addition of an expensive third laser in the blue region is not essential: the green laser provides the luminosity to the hologram while its combination with the red laser gives a wide range of colour shades. Holograms made with red and green lasers offer a good compromise between cost and effect on the audience. As we will see later in the text, the use of two red/green lasers has also the advantage to enrich considerably the possible experiments for teaching the principles of holography. Bigger holograms can be realized with the kit and a different plate/object support adapted to the plates size (until 8"x10"), but experiment is much more sensitive to vibrations.

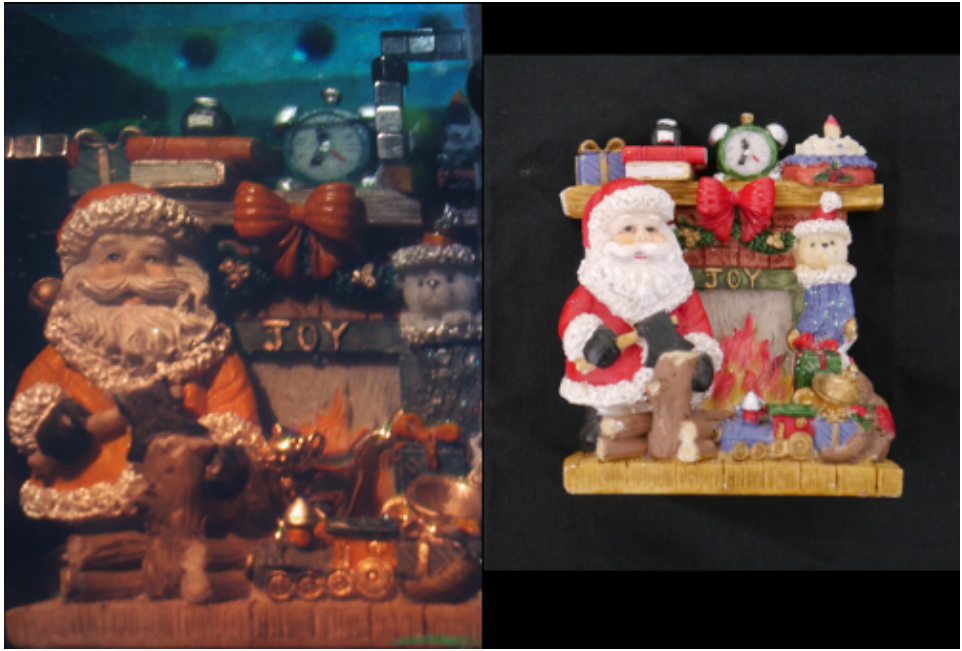

**Figure 3** Two-colours (red/green) hologram of a Christmas toy (left), and the original object (right).

### 3.3. Single shot transmission/reflection holograms

There are two main categories of holograms. First, transmission holograms were invented by D. Gabor in 1948 [14]. He presented an experiment with a low coherence reference beam whose wavefront was distorted by passing through a transparent object (object beam), creating an interference pattern that could be recorded in the form of a hologram. But light sources with a good temporal coherence did not exist at this time. This greatly limited the development of holography after D. Gabor's experiment. The invention of lasers in the sixties [15], provided light sources with a much better coherence than classical sources, and then an easy way of creating holograms. A transmission hologram set-up is characterized by the relative disposition of the object and reference beams: they must come from the same side of the light-sensitive medium. In this configuration, the global orientation of the interference pattern is perpendicular to the plate surface. Once developed, these holograms are seen by looking at the reference beam through the holographic plate. The second kind of holograms was invented in 1962 by Y. Denisyuk [12] and can be seen with the reflection of a white light spot. Such holograms are obtained when the object and reference beams come from the opposite sides of the light-sensitive medium. In this configuration, the global orientation of the interference pattern is parallel to the plate surface, which gives the hologram the ability to filter the incoming white light [10]. When teaching practical holography, it is important to explain the difference between these two main kinds of holograms. In the frame of the pedagogic tool for colour holography, the plate/object support PO2 has been specifically developed for this purpose. Indeed, it allows realizing at the same time a transmission and a reflection hologram by placing a first object behind the holographic plate (single beam reflection hologram), and a second object in front of the plate (single beam transmission hologram). Once developed, such a hologram has the particularity to contain in the same volume a reflection and a transmission hologram. Depending on the way of observing it with the laser by transmission or with a white light by reflection, the student can see each of them. This experiment has a great pedagogic value: it enables the student to question the effect of the relative orientation of the object/reference beams with respect to the plate.

The Bragg condition determines the observed colour of a reflection hologram lightened with white light [16]. Indeed, only light of wavelength equal to twice the fringe spacing is reflected from the fringe planes in phase to give a constructive interference and consequently, a bright image. Holographic plates can be pre-swelled, which means that when developed, the emulsion width decreases, leading to a lower fringe spacing in the case of a reflection hologram, and to a colour shift of the image towards the small wavelengths. To make students understand this particular property of reflection holograms, it is interesting to make single shot transmission/reflection holograms with the green laser of the kit and pre-swelled plates. Once developed, the student will observe a green transmission image and a blue-shifted colour reflection image. He will need to understand the concept of the Bragg effect to explain why the colours of the two images are different.

*3.4. Holographic diffraction grating*

Holographic diffraction gratings are widely used in spectroscopy for research and industrial spectroscopic applications. Such a optical element is obtained when two coherent equally polarized monochromatic optical beams of equal intensity intersect each other onto a photo-sensitive plate. A diffraction grating can also be created from a single collimated coherent beam, if it is reflected back upon itself. The use of the plate/object support PO1 allows this configuration by placing the mirror vertically against the rear post, with the holographic plate placed at the original mirror position (figure 4) [17]. A standing wave pattern will be formed, with intensity maxima forming planes parallel to the wavefronts. Its intersection with a photo-sensitive medium yields on its surface a sinusoidally varying intensity pattern, whose spacing $d$ depends on the angle θ as defined on figure 4. For a collimated beam and a laser wavelength λ, we have:

$$d = \frac{\lambda}{2 \sin \theta} \qquad (1)$$

The number of grooves per millimeter $1/d$ can be controlled by choosing the laser wavelength λ and the angle θ. With the pedagogic case for colour holography, it is possible to obtain values up to 2800 mm$^{-1}$ on a 4"x5" holographic plate.

When realizing diffraction gratings with the kit, the laser beam is divergent. Then, the optical wavefront is spherical and relation (1) is not strictly valid. This implies that the distance between successive fringe patterns 1) differs from the value for a collimated beam and 2) is not constant over the diffraction grating surface. To check how important these effects are, we measured that the variation of the value of the number of grooves per millimeter over the surface of a diffraction grating realized with the kit was less than 4 %. Moreover, we realized a diffraction grating with a collimated beam by adding a second converging lens after the one used to expand the laser. In that case, the resulting number of grooves per millimeter was 2.2 % lower than that of the diverging beam configuration. Then, we can conclude that the relation (1) is a good approximation in our case. These results can be understood if we consider that the plate/object support PO1 is positioned far from the laser sources, at a distance L≈4.5 m, so that the incoming beam is only slightly spherical.

Once realized, the number of grooves per millimeter $a_{\exp}$ can be experimentally obtained by measuring the angle of diffraction $β_2$ of an incident collimated laser beam with an incident angle $β_1$ with respect to the normal [18]:

$$\sin(\beta_1) + \sin(\beta_2) = \frac{\lambda}{d_{exp}} = \lambda \cdot a_{\exp} \qquad (2)$$

For example, a diffraction grating was made with the green laser and the mirror inclined with an angle θ = 41.5°. The use of relation (2) with two lasers at 671 nm and 473 nm gives $a_{\exp}$ = 2513 mm$^{-1}$ and 2457 mm$^{-1}$, respectively. This is in very good agreement with the value predicted from equation (1): $a_{th}$ = 2491 mm$^{-1}$. One can also study the effect of the wavelength on the resulting value of $d$. Two diffraction gratings were successively made with the red and green lasers of the kit. The use of relation (2) gives $a_{exp\ 632}$ = 2254 mm$^{-1}$ and $a_{exp\ 532}$ = 2714 mm$^{-1}$. As expected, the ratio $a_{exp\ 532}$ / $a_{exp\ 632nm}$ = 1.19 is equal to the ratio of the laser wavelengths $\lambda_{red}$ / $\lambda_{green}$. The realization of these two gratings can be made at the same time on the same plate, so that two holographic gratings can coexist with two +1 orders and the corresponding resolutions.

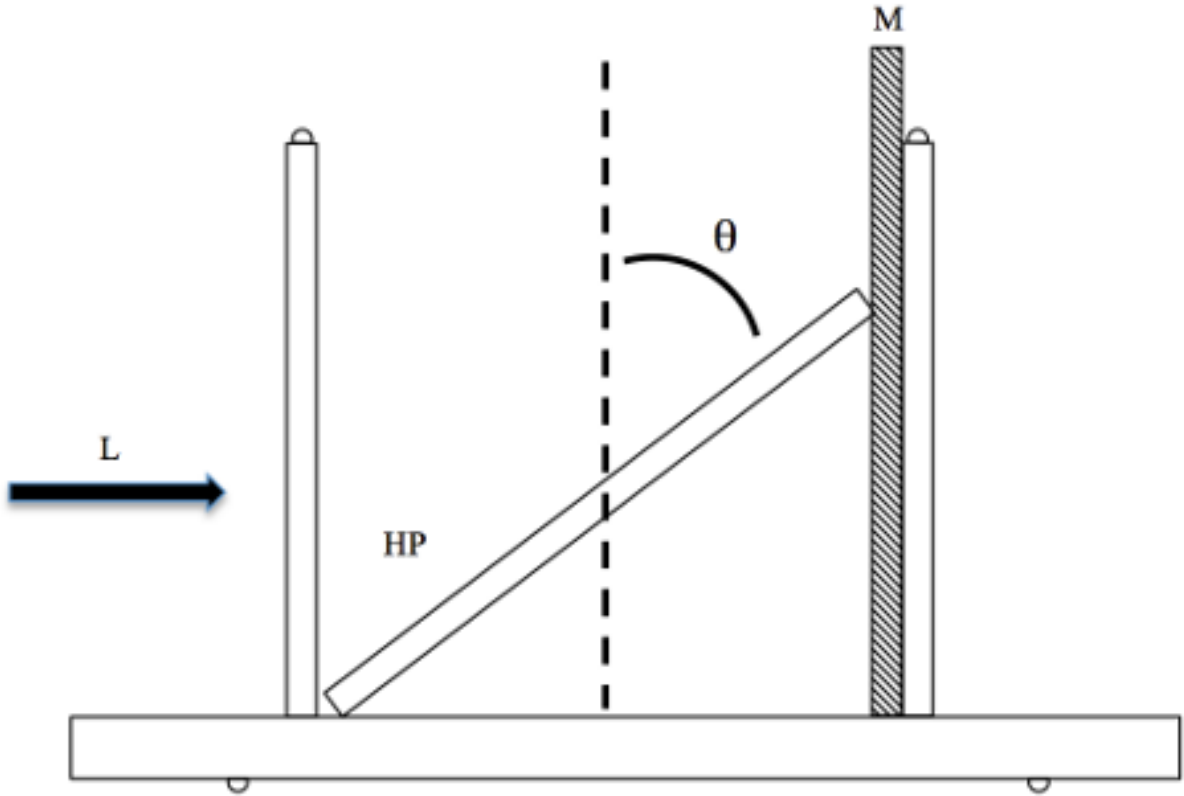

**Figure 4** Experimental setup for the realization of a holographic diffraction grating with the plate/object support PO1. The laser beam L is coming from the left side. The holographic plate HP is leaned against the mirror M, emulsion on the mirror side. The inclination of HP can be varied by changing the position of the rear post.

*3.5. Wavelength and angular multiplexing*

Information storage using discs (DVD, blu-ray, etc) are based on a surface technology, by simply making microscopic holes on a plastic disc to write down binary informations. These technologies are now well known and their capacities close to their theoretical limits. Holographic versatile discs (HVDs) technology allows storage of data in volume, which greatly increases the density of information. Moreover, angular and wavelength multiplexing can also be used to further increase it. This technique relies on some singular properties of holography: by changing the reference beam angle or the laser wavelength between several exposures of the same medium, it is possible to record several sets of data on the same photo-sensitive volume. After development, the hologram returns a particular set of data depending on the reference beam angle and laser wavelength used to read the disc. Our holography kit enables illustrating these multiplexing principles. For angular multiplexing, the plate/object support PO2 is simply rotated to change the incident angle of the laser beam between two laser exposures of two different objects on the same plate. Once developed, one can see the image of one object or the other, depending on the angle of observation of the hologram. This experiment can be done with a transmission or a reflection hologram, depending on the position of the object in front of or behind the plate. The wavelength multiplexing effect can be realized with the PO1 plate/object support: the same holographic plate is successively exposed to the two lasers, changing the object between the shots. Once developed, the two red and green images are seen superimposed when the hologram is lit with white light. When it is exposed to red and green lights, one can see the corresponding image, thus illustrating the wavelength selectivity of the hologram.

*3.6. Example of use of the pedagogic kit for colour holography*

The colour kit was first tested during a French popular science event, the science day, in October 2015 in Marseilles. Aided by two Master students, we realized during one day about twenty beautiful 4"x5" colour holograms for the enjoyment of people of all ages. The environment was very noisy, but it did not influence the quality of the holograms we made. The colour kit was also used for colour holograms realization during several one-hour meetings with junior-high and high school students, and during the presentation of a typical first year course for high school students interested in discovering the university. In the frame of a project on the different kind of 3D images, four groups of high school students came to the university holography lab to make colour holograms with the kit. I also gave several conferences on holography in France, ending my presentation by the realization of a beautiful colour hologram. The colour kit was also used in complement to the monochromatic kit for a one day continuing education formation of junior-high and high school teachers. A second version of the colour kit is being built for a popular science center in France.

The experiments described in this work enables to present fundamental and applied holography to university students. The kit has several advantages in comparison to classical experiments: it is autonomous, portable and does not imply any expensive anti-vibration device. It provides a wide range of experiments for 3 h sessions or longer projects. For example, the colour kit was used for experimental projects with small groups of Master students of the Aix-Marseilles University. During 6 sessions of 3 hours, they first worked on the realization and the qualification of diffraction gratings. They began by making a single shot transmission/reflection hologram to understand the differences between them and to master the different phases of realization of a hologram: setup mounting, exposure time calculation, and development. Then, they made and qualified several diffraction gratings and mounted a complete transmission spectrometer with it [19].

**4. Conclusion**

Our first monochromatic teaching kit for holography developed in 2012 has been widely used for teaching purposes at the Aix-Marseille University, for continuing education and for popular science purposes. Since 2015, it has been upgraded to a colour version described in this publication. It provides more pedagogic subjects for university students with no particular need for an expensive active or passive anti-vibration device. With the new colour kit, it is possible to develop more complex educational experiments, complementary to the ones proposed with the monochromatic kit: colour holography, comparison of reflection/transmission holograms, Bragg effect, angular and wavelength multiplexing principles, diffraction gratings with control of the number of grooves per millimeters, notch filters with central wavelength control. Moreover, the kit is portable and can also be used for outreach purposes, during public exhibitions, science festivals, conferences, and for junior-high and high-school teacher training. Holography has the advantage of allowing a deep study of a wide range of fundamental principles of optics. At the same time, the incredible ability of holography to reproduce reality acts as an excellent way of motivation for students. It is also a very efficient manner to capture the attention of an audience, whoever it is.

**Acknowledgments**

The authors would like to thank Yves Gentet for helpful discussions. This work was supported by the French government's facility "Science à l'école", as part of the LUNAP operation.

**References**

[1] Th. Voslion and A. Escarguel, *Eur. J. Phys.* **33,** 1803-1811 (2012)

[2] A. Escarguel, *Journal sur l'enseignement des sciences et technologies de l'information et des systèmes*, **9**, 0015 (2010)

[3] A. Escarguel, *proceedings of the congrès général de la Société Française d'Optique*, Paris, France (2013)

[4] Th. Voslion, A. Escarguel, International Symposium on Display Holography, MIT Media Lab, Cambridge Massachusetts USA, *J. Phys.: Conf. Ser.* **415**, 012001 (2012)

[5] A. Escarguel, *Colloque sur l'Enseignement des Technologies et des Sciences de l'Information et des Système*, Trois-rivières, Québec (2011)

[6] A. Escarguel, *rencontres pédagogiques, congrès "Optique Marseille" (COLOQ, HORIZONS, JNCO et JNOG)* (2011)

[7] A. Escarguel, *Colloque sur l'Enseignement des Technologies et des Sciences de l'Information et des Système*, Grenoble, France (2010)

[8] A. Escarguel, *proceedings of the 10$^{th}$ International Symposium on Display Holography*, St Petersbourg, Russian federation, (2015)

[9] A. Escarguel and R. Baude, invited talk, *rencontres pédagogiques du congrès général de la Société Française d'Optique*, Bordeaux, France (2016).

[10] G. Saxby and S. Zacharovas, *Practical holography* (4th edition, CRC Press, New York, 2015)

[11] http://www.ultimate-holography.com/

[12] Y. N. Denisyuk, *Akad. Nauk SSR* **144** 1275–8 (1962)

[13] K. Bazargan, *International Symposium on Display Holography*, MIT Media Lab, Cambridge Massachusetts USA, J. Phys.: Conf. Ser. 415 012028 (2012)

[14] D. Gabor, *A new microscopic principle*, *Nature* **161,** 777 (1948)

[15] T.H. Maiman, *Nature*, **187**, 493 (1960)

[16] H. J. Caulfield, *Handbook of optical holography,* (Academic Press, New York, 1979)

[17] N. K. Sheridon, *Appl. Phys. Lett.* **12**, 316 (1968)

[18] J. James, *Spectrograph design fundamentals (*Cambridge University Press, Cambridge, 2007)

[19] R. E. Bell, *Rev. Sci. Instrum.* **75**, 4158 (2004)